# Self-Propelled Janus Colloids in Shear Flow

Bishwa Ranjan Si[a], Preet Patel[a] and Rahul Mangal[a*]

[a] Department of Chemical Engineering, Indian Institute of Technology Kanpur

To fully harness the potential of artificial active colloids, investigation of their response to various external stimuli including external flow is of great interest. Therefore, in this study, we perform experiments on $SiO_2$-*Pt* Janus particles suspended in an aqueous medium in a capillary subjected to different shear flow rates. Particles were propelled using varied $H_2O_2$ (fuel) concentrations. For a particular propulsion speed, with increasing shear flow, a transition of motion of active Janus particles from the usual random active motion to preferential motion across stream-lines and then finally to migration along the flow, was observed. Our analysis revealed that these transitions are dictated by the torque due to the self-propulsion near wall w.r.t. shear-induced torque. Interestingly, we found that only when these torques are comparable, particles align in a manner such that they migrate significantly across the streamlines.

## INTRODUCTION

Transport of colloidal particles (1 nm - 1000 nm) is an underlying process of many technologies such as paints, separation processes, drug delivery etc. Thermal fluctuations induce Brownian motion which offers poor control over the speed and directionality of colloidal particles. Application of external fields, such as electrical or magnetic fields, has been successfully utilized for gaining some control over the transport of colloids.[1] The disadvantage, however, is that application of external field results in collective convective motion of particles and moreover, poses the hazard of affecting the surroundings unpredictably. Inspired from the microscopic biological systems such as bacteria and unicellular protozoa which use their flagella or cilia for active swimming, [2,3] and from the macroscopic phenomena such as flocking of birds and moving school of fish,[4–6] recently it has been realized that synthetic colloidal particles (~ 1-1000 nm) can also harness the energy from the environment. Under appropriate conditions, local gradients/fields are created at the individual particle level, resulting in net forces which propel these colloids, individually, out of their equilibrium. Such particles are hence known as self-propelled or active particles. Typically, Janus particles (JPs)[7,8] are widely employed for active propulsion due to their in-built asymmetry generating chemical potential. The term "Janus" was coined from the Roman God Janus, with two heads placed back to back, one looking to the future and other to the past. The context here being JPs have asymmetric faces with completely different properties, i.e. one face is reactive while other is non-reactive. In the most commonly used mechanism to achieve active propulsion one half of the JP is a dielectric material such as $SiO_2$, $TiO_2$ or Polystyrene and the other half is coated with either metal Platinum or Palladium. These JPs are then introduced into $H_2O_2$ aqueous solution.[9–11] The catalytic nature of the metal site promotes the decomposition of $H_2O_2$, and the resulting concentration gradient of $O_2$ molecules leads to self-diffusiophoresis[12,13] of the particle. However, the mechanistic details continue to be a subject of discussion.[14] Other potential mechanisms for self-propulsion include laser-induced self-diffusiophoresis in a phase separating mixture,[15] and self-thermophoresis.[16] Unlike passive colloids which undergo random Brownian motion, active colloids feature various novel transport and self-assembly behaviours. Researchers believe that these unique features of synthetic active colloids, apart from being fundamentally exciting, can potentially be used in a multitude of applications including drug delivery, macromolecular separations and colloidal self-assembly. [17–19]

For most of the sought-after applications the surrounding media of the active particles is generally a complex fluid that contains particulates and/or biopolymers. The particles may also experience a local shear environment induced by external flow. However, so far, the motion of synthetic active particles, has been experimentally investigated mostly in the bulk [9,10,16,17,20,21] or interfaces[22–24] of "quiescent" fluids. Only few studies have focused on investigating the effect of physical obstacles.[25–28] Most of the work on understanding the active motion in shear flow focuses on living microorganisms.[29,30] For synthetic active colloids, very few experimental or analytical studies have been performed. In 2015 Palacci *et al.*[31] reported positive rheotaxis in light-activated haematite cubes embedded in a polymeric sphere. Similar rheotactic behaviour has also been studied for bimetallic rods.[32,33] Using mesoscale simulations, Shalabh and Sunil[34] recently predicted a combination of upstream and downstream rheotaxis for self-propelled filaments. In all these cases, the anisotropy in shape was found to be critical in governing the particle motion due to the "weather vane" mechanism. For symmetric spherical active particles combining analytical and numerical calculations, Uspal *et al.*[35] suggested rheotactic behavior via a mechanism involving ''self-trapping'' near the hard wall. Very recently, Katuri *et al.*[36] conducted experiments using a couple of combinations of activity and shear flow rate. They demonstrated, that on subjecting the spherical active JPs to external shear flow particles drift perpendicular to the flow, which is in sharp contrast to the previous simulation predictions.

Considering both the scarcity of relevant studies and the consensus among the existing ones, further studies are required to gain an in-depth knowledge of the effect of shear flow on the motion of spherical active JPs near a planar boundary. Therefore, in this work we perform detailed experiments to obtain further insights. Our experiments will explore this effect over a wide range of relative strengths of activity and external shear flow rates.

## EXPERIMENTAL

### Synthesis of Janus particles

Janus particles were synthesized using the drop casting procedure reported by Love *et al.*.[37] Following the protocol, a monolayer of isotropic $SiO_2$ (~ 5 µm) colloidal particles was prepared by drop casting the dilute suspension of colloids (in Ultrapure water) on a clean oxygen-plasma treated glass slide. The plasma treatment makes glass slide hydrophilic, resulting in uniform spreading of the particle suspension. Water was subsequently removed by slow evaporation leading to the formation of a monolayer which was confirmed via optical microscopy. After the formation of monolayers, a thin layer of Pt (~15-20 nm) was deposited from the top using plasma sputtering deposition in a sputter coater. Due to self-shadowing effects, one hemisphere of the particle gets coated by the Platinum metal, and thus JPs are obtained. The SEM micrograph, as shown in figure 1(b), confirms the metal coating. After sonication for few minutes JPs were dispersed into deionized water.

### Shear-flow set-up

To investigate the motion of active particle in flow, dilute suspension of JPs was introduced into a piranha cleaned glass capillary (Vitrocom) of square cross section (1 mm x 1mm), connected to a 1 ml glass syringe mounted on a syringe pump (Chemyx 2000). Particle solution in the capillary was subjected to different volumetric flow rates (i.e. 0, 100, 200, 500, 750, 900, 1000 µL $hr^{-1}$), along the length of the capillary towards -X direction (see figure 1(c)).

### Particle Tracking

Since the particles are heavy, they sediment to the bottom of the capillary and perform 2D motion (X(µm)-Y(µm)) by hovering above the bottom wall. All the particles were observed within the same focal plane suggesting that they remain at similar vertical height from the bottom wall. Using the syringe pump a pressure driven flow was generated in the capillary. However, since the particles are near the bottom wall and are much smaller in size compared to the capillary diameter, the local flow profile for the particles can be assumed as simple shear flow. An Olympus LC30 camera with 2048 x 1532 pixel resolution, equipped with Olympus upright optical microscope

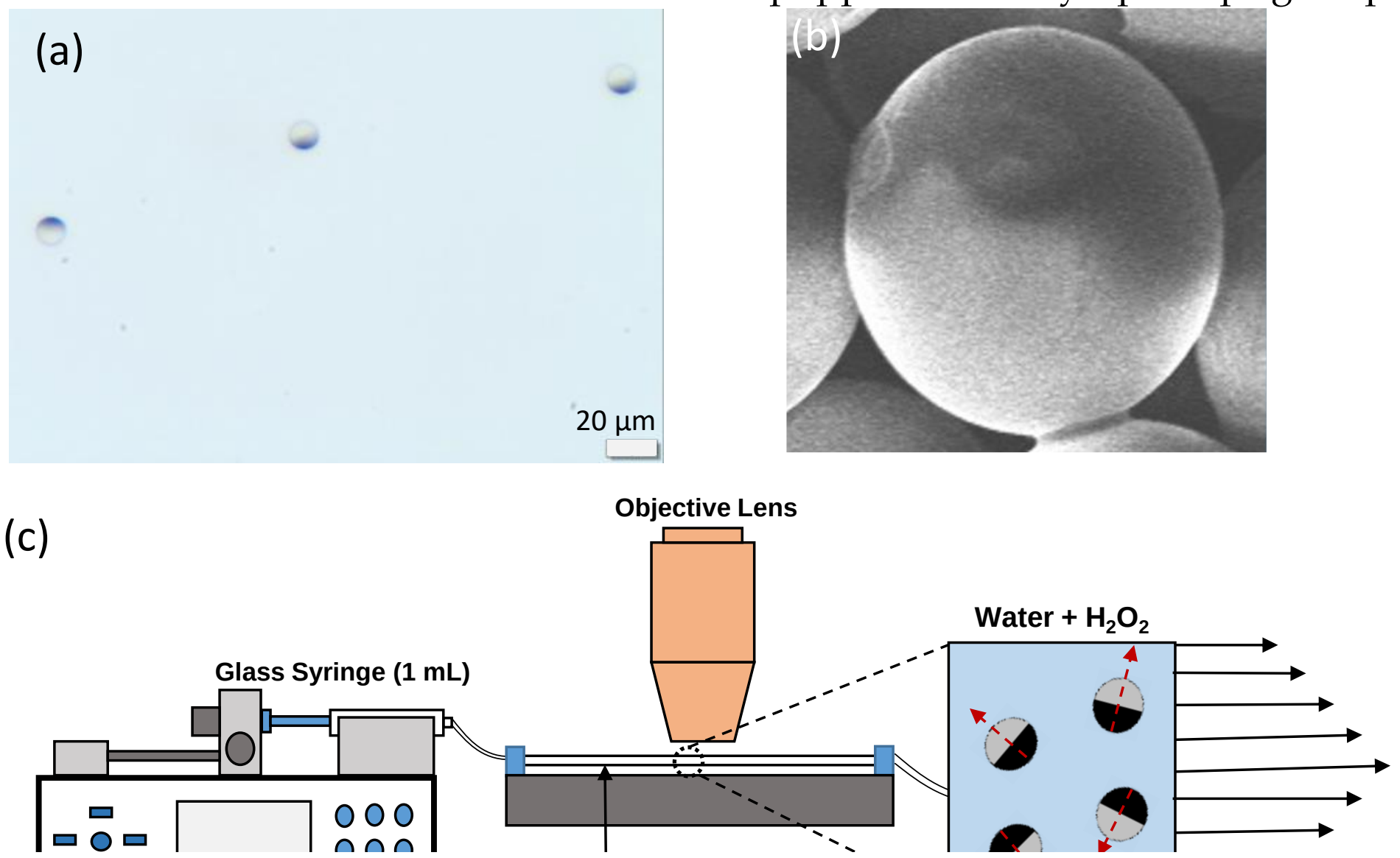

Figure 1: Figure 1: (a) Bright-field micrograph of $SiO_2$-*Pt* JPs (b) SEM micrograph of $SiO_2$-*Pt* JP (c) Schematic of the experimental setup.

BX53 was used to image the JPs. In order to ensure negligible effect of the vertical side walls of the capillary on our measurements, we focused within a narrow Y range (~ ±100 μm) along the center line of the capillary. Motion of isolated JPs, in the X-Y plane, were recorded for 5-10 mins at 22 fps. Particle tracking was performed with Image-J using an image correlation-based approach to obtain particle trajectories (X(μm), Y(μm) vs. $\Delta t$). All the trajectories were analyzed for similar durations ~ 40-50s. A calculated amount of 30 wt % $H_2O_2$ aqueous solution was added to particle solution to bring the final concentration of $H_2O_2$ in the particle solution to the desired value (0, 0.5, 1, 2 wt %) to vary the propulsion speed.

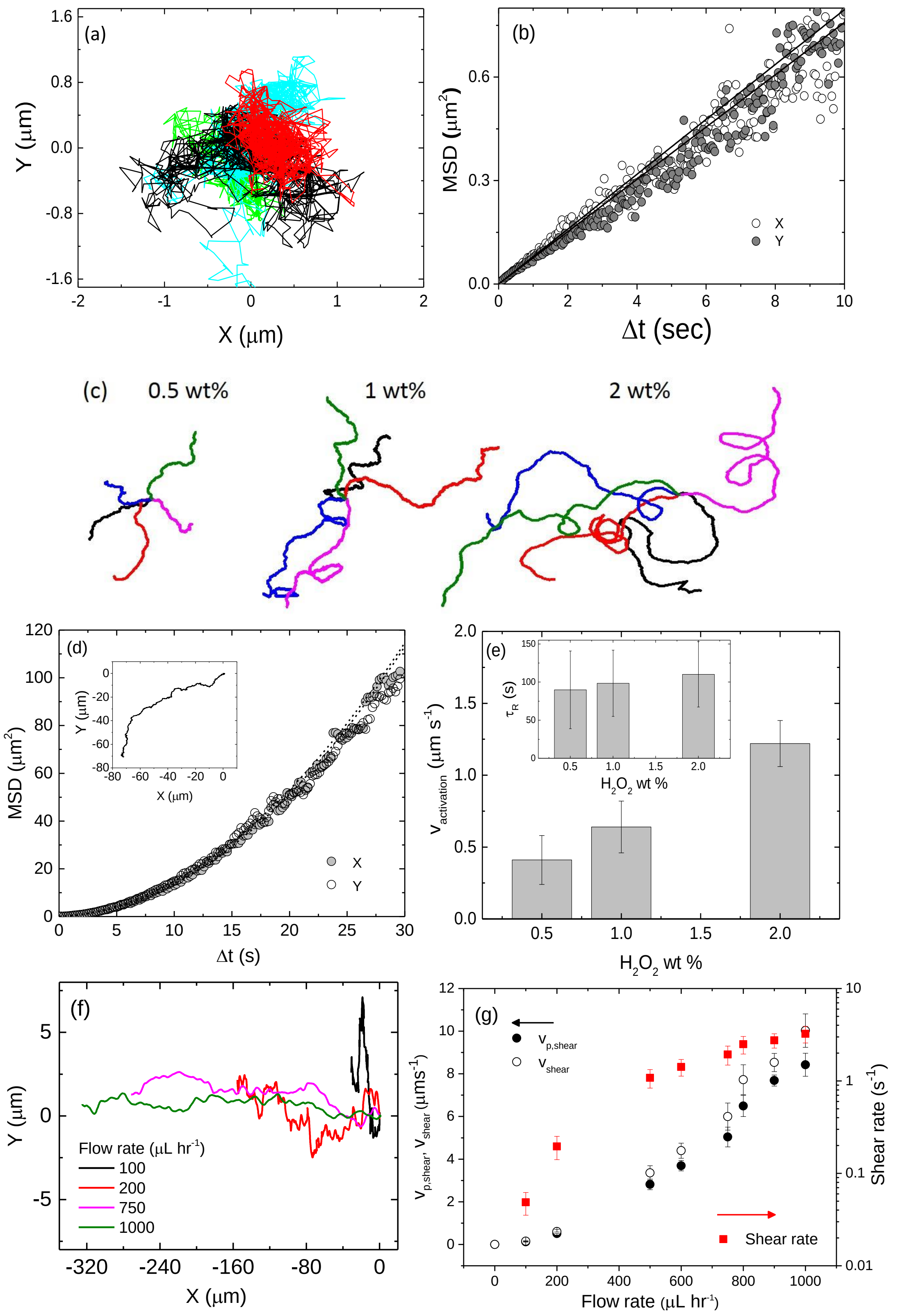

Figure 2: (a) X-Y trajectories of Brownian JPs (b) MSD vs time for a selected Brownian trajectory (c) Trajectories of active Janus particles (d) MSD of an active particle with 2 wt% $H_2O_2$. Inset shows the corresponding X-Y trajectory (e) Variation in propulsion speed v as a function of $H_2O_2$ wt %. Inset shows the rotational time scale ($\tau_R$) (f) Trajectories of inactive Janus particles in external flow (g) Variation in $v_{p,shear}$, $v_{shear}$ and shear rate ($\gamma$) of the inactive JPs with flow rate.

## RESULTS AND DISCUSSION

### Brownian motion

Firstly, to validate our experimental set-up, we performed experiments to investigate the 2D *Brownian* motion of the JPs in the absence of $H_2O_2$ and the external flow. In Figure 2(a), we show the tracked *X-Y* trajectories for a few particles. Mean square displacement MSD (X,Y) as a function of $\Delta t$, for one of the trajectories is shown in figure 2(b). Consistent with Brownian motion, MSD(X) and MSD(Y) vary linearly with $\Delta t$. Furthermore, absence of any anisotropy in the MSD values confirmed the absence of any external convection. 2D diffusivity values were computed from the slope of the MSD (X,Y) vs $\Delta t$ resulting in $D(X) \sim D(Y) \sim 0.04 \pm 0.01\ \mu m^2 s^{-1}$. Note that the obtained values are smaller than the theoretical estimate of 0.09 $\mu m^2 s^{-1}$ obtained using Stokes-Einstein equation $D_{SE} = kT / 6\pi\eta a$, where T = 298 K, $\eta$ (~ 8.9 X $10^{-4}$ Pa-s) is the bulk viscosity of water and $a$ (~ 2.5 μm) is the particle radius. This reduced or hindered tangential $D_{||}$ (D(X) or D(Y)) is because of the interaction of the JP with the bottom wall, which is consistent with the near wall hindrance theory.[38]

### Active motion in absence of flow

Next, we analysed the active motion of JPs in the presence of $H_2O_2$ but in the absence of flow (see supporting information for sample video). Figure 2(c) shows the X-Y trajectories for different $H_2O_2$ concentrations. As expected, at higher $H_2O_2$ concentrations, within the same time frame particles travel larger distances. Further, from the trajectories we calculated the MSD (X,Y) vs $\Delta t$. In figure 2(d), we show the MSD for an active particle exposed to 2wt% $H_2O_2$. Parabolic nature of the trajectories confirms the activity of particles and lack of anisotropy in the MSD in the orthogonal directions confirm the absence of bulk convection. The motion of an active JP is subjected to a change in direction which is controlled by its characteristic rotational time $\tau_R^{-1} = kT / 8\pi\eta a^3$.

For short times i.e. $\Delta t << \tau_R$, MSD (X,Y) motion is primarily ballistic and can be fitted using the following equation

$$\Delta L^2 = 4D_o \Delta t + \mathrm{v}_{\mathrm{activity}}^2 \Delta t^2 \quad \ldots\ldots\ldots (1)$$

Here, $\Delta L^2$ is the 2-D MSD (X,Y), $D_o$ is the Brownian diffusivity and $\mathrm{v}_{\mathrm{activity}}$ is the propulsion speed of the particle. Figure 2(e) shows thus obtained values of $\mathrm{v}_{\mathrm{activity}}$ using the fit. An increasing trend in $\mathrm{v}_{\mathrm{activity}}$ with increasing $H_2O_2$ is expected and is in agreement with the previous studies.[10] Velocity autocorrelation was used to compute $\tau_R$ using the following equation, supporting figure 1S:

$$C(\Delta t) = 4\left( D_o / \mathrm{v}_{\mathrm{activity}}^2 \right)\delta(\Delta t) + \exp^{-2\Delta t/\tau_R} \quad \ldots\ldots\ldots\ldots\ldots (2)$$

Here, the first term captures the sharp decay in correlation at very short times caused by equilibrium Brownian fluctuations. The second term captures the decay over the long times, at the rotational time scale. As shown in the inset to figure 2(e), increasing $H_2O_2$ does not affect the $\tau_R$ values significantly. Lack of any correlation between the increase in translation speed/fuel and the rotational time, in a stagnant Newtonian medium is in line with existing literature.[10] Additionally, we noticed that most active JPs oriented with their Janus plane perpendicular to the wall surface i.e. 'half-moon orientation' (see figure 2S). This observation is consistent with previous reports[26,39] and has been discussed in detail in later section.

### Inactive motion in flow

We also analysed the trajectories of inactive Janus particles in an imposed shear flow (–X direction) at different volumetric flow rates (see supplementary information for sample video). In figure 2(f), few representative trajectories at different flow rates have been shown. It is clear that, at low flow rates due to the significant contribution of Brownian fluctuations, the drift of Janus particles in Y direction in comparison to drift along -X direction, is not negligible. However, at higher flow rates, the relative drift in Y direction decreases. Using the drift in -X direction, we calculated the -X velocity of the Janus particles $\mathrm{v}_{\mathrm{p,shear}}$ and observed monotonically increasing trend with increasing flow rates (see figure 2(g)) which is expected. To compute the shear flow velocity ($\mathrm{v}_{\mathrm{shear}}$) and the corresponding shear rate ($\dot{\gamma} = \mathrm{v}_{\mathrm{shear}}/a$) at the particle length scale, we need to know the height ($h$) of the particles' center from the bottom wall, which is not directly measurable. Hence, we estimate it using the near wall hindrance theory described by Kihm *et al.*[38]. The study uses the classical Faxen's expression (1923) to compute $h/a$ from the ratio of the experimentally measured Brownian diffusivity parallel to the wall ($D_{||}$) with respect to the expected bulk Stokes-Einstein Diffusivity ($D_{SE}$) i.e. $D_{||}/D_{SE}$. Further, using $h/a$ and $\mathrm{v}_{\mathrm{p,shear}}$ we calculated $\mathrm{v}_{\mathrm{shear}}$ (see Goldman *et al.*)[40]. In figure 2(g), we show the values for thus calculated $\mathrm{v}_{\mathrm{shear}}$ and $\dot{\gamma}$, and experimentally measured $\mathrm{v}_{\mathrm{p,shear}}$. Clearly, $\mathrm{v}_{\mathrm{p,shear}} \approx \mathrm{v}_{\mathrm{shear}}$ and our chosen flow rates cover two orders of magnitude of $\dot{\gamma}$ from ~ $0.05 s^{-1}$ to $5 s^{-1}$. Additionally, we would also like to report that for the highest flow rate, the Reynolds number at the particle length scale is ~ $10^{-5}$. This confirms the absence of any flow inertial effects at the relevant length scales.

## Active motion in flow

Finally, we analyzed the motion of JPs with different activities exposed to a range of different shear flow rates (see supporting information for sample videos). In figure 3, we show a few representative X-Y trajectories of active JPs. Few of the captured videos are available in the supporting material (videos S1-S4). Based on simple inspection of the trajectories, we notice that at very low flow rates (~0-100 μL $hr^{-1}$), JPs perform random active motion at long times and the trajectories are near isotropic in nature, confirming minimal effect of shear flow. At intermediate flows, we observe that the dynamics of active particles is noticeably affected by the shear flow and the trajectories start to become anisotropic in nature. Unexpectedly, most trajectories were elongated along Y, suggesting preferential migration of JPs perpendicular to the flow direction. At higher flow rates this tendency of particles to migrate across the flow decreases and trajectories become more elongated along X. However, their Y drift was found to be significantly larger than their inactive counterparts at the same flow rate. We also noticed that the tendency of the particles to migrate along the flow direction was achieved at progressively higher flow rates with increasing activity. To understand these observations more carefully and to quantify the effect of flow on the trajectories of active JPs, we computed a non-dimensional parameter ($\alpha = \dfrac{\left( \Delta Y_{max} / \Delta X_{max} \right)_{activity+flow}}{\left( \Delta Y_{max} / \Delta X_{max} \right)_{flow}}$), which measures the anisotropy along Y, in the trajectory of *active* JPs in flow, relative to the anisotropy in trajectories for the *inactive* JPs in flow. The two-step normalization in this definition allows us to compare the anisotropy in trajectories with minimal bias to JPs' initial X position as well as the flow-rate. $\alpha \sim 1$ indicates absence of any activity induced anisotropy in the trajectory in the presence of flow.

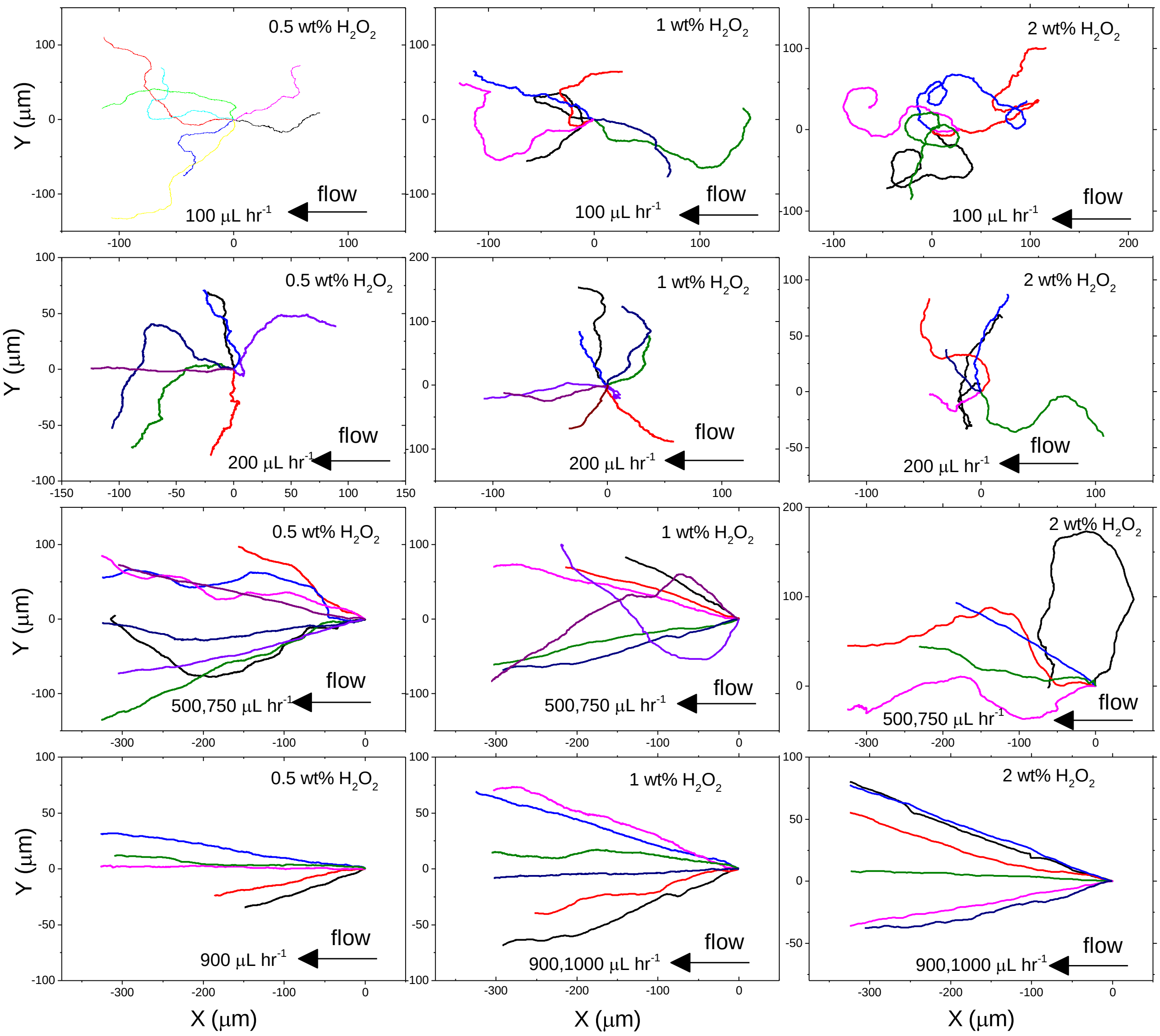

Figure 3: Trajectories of active Janus particles in external shear flow.

On the other hand, $\alpha >> 1$, suggests that activity coupled with flow induces preferential migration of JPs in Y direction. In figure 4, we plot the distribution of $\alpha$, for around 20-40 JPs, at selective flow rates ($v_{shear}$) for 0.5wt% and 1wt% $H_2O_2$. Additional data for remaining flow rates and for 2wt% $H_2O_2$ is available in supporting figure 3S-5S. Clearly, for all $H_2O_2$ concentrations, in the absence of flow, $\alpha$ is distributed close to 1 due to the lack of any activity induced anisotropy in the motion.

With increasing flow-rate, we notice that both the range of $\alpha$ and the population of particles with higher $\alpha$ increases, suggesting that particles prefer to migrate along Y in flow due to active propulsion. At very high flow rates, the distribution of $\alpha$ again starts to shrink back towards lower values, suggesting that lesser particles migrate along Y. Interestingly, with increasing $H_2O_2$ concentrations, the transition of $\alpha$ towards lower values is achieved at progressively higher flow rates. This prompted us to investigate the variation in $\alpha$ using a non-dimensional number $\lambda = \frac{|v_{p,shear}|}{|v_{activity}|}$, which measures the strength of particle speed due to shear flow relative to the its propulsion speed. According to figure 4, the variation in $\alpha$ follows a universal trend w.r.t. $\lambda$, for any $H_2O_2$ concentration and flow-rate. For low $\lambda$ (< 0.5) where propulsion speed is much stronger compared to the imposed shear, particles' movement appear to be nearly unaffected by the flow and hence $\alpha \sim 1$. On the other extreme i.e. for large $\lambda$ (> 5), where the imposed shear flow is much stronger than propulsion speed, particles appear to be dragged along the flow, which again results in $\alpha$ values close to 1. At moderate $\lambda$ (0.5-5), where both shear flow and propulsion speed are comparable, we observe maximum extent of migration across the flow.

We also observed that, at higher flow rates trajectories are nearly rectilinear with very less in-plane (about Z axis) rotation. This is not the case at low to moderate flow rates where particles frequently changed their in-plane direction of motion. To quantify this observation using the velocity auto-correlation function, we computed the rotational diffusion coefficient $D_r$ ($=1/\tau_R$) of JPs at different $\lambda$, supporting figure 6S. We notice a shift in the distribution of $D_r$ to lower values that with increasing $\lambda$. For smaller $\lambda$, since the effect of imposed shear is weak active JPs frequently change in-plane orientation which is an expected behavior.[9,10] With increasing $\lambda$, the strong imposed shear flow forces the particles to rotate out-of-plane i.e. about Y axis and does not allow it to change the in-plane direction of motion, thereby reducing $D_r$.

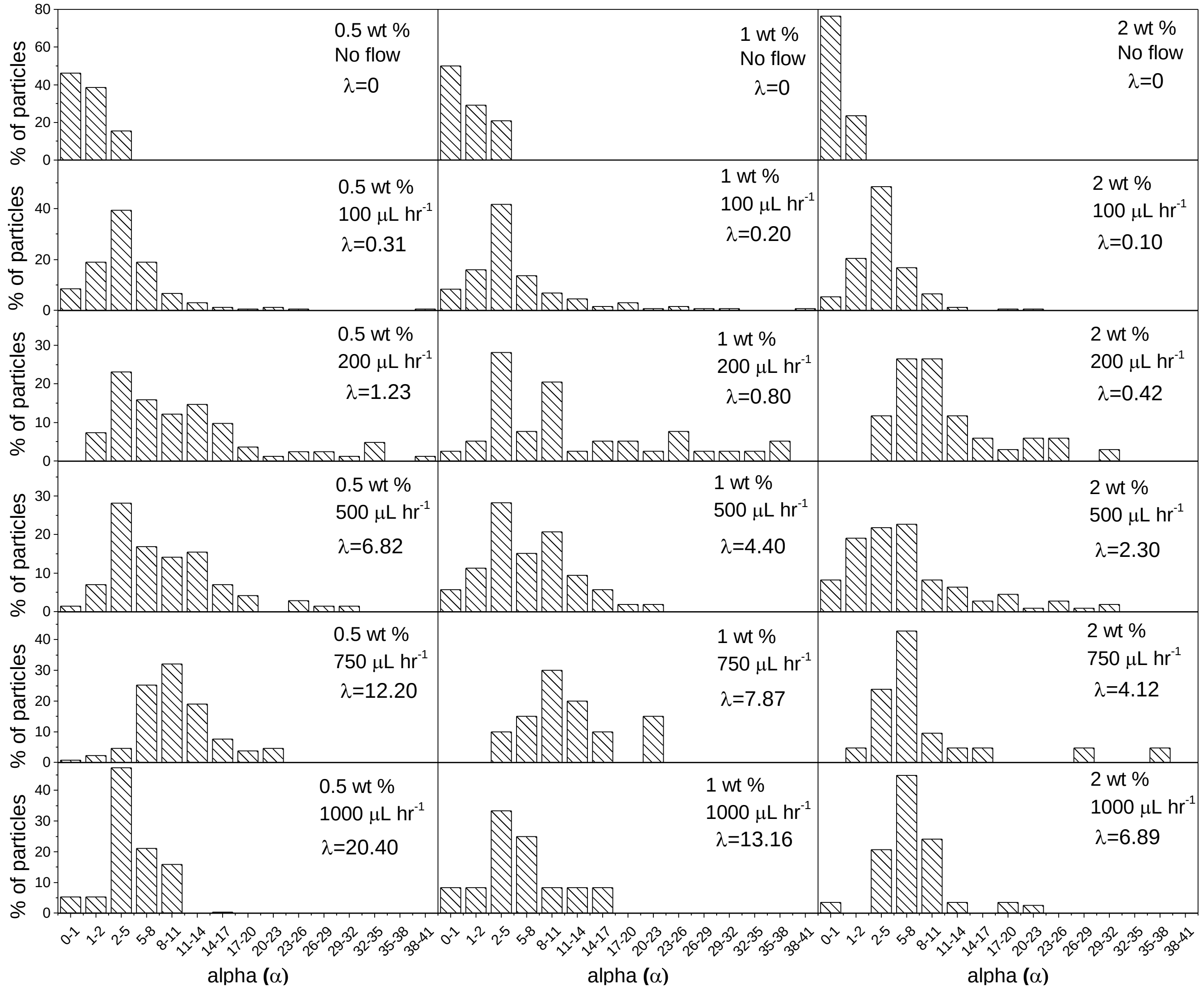

Figure 4: Distribution of $\alpha$ with varying $H_2O_2$ concentrations and varying flow rates.

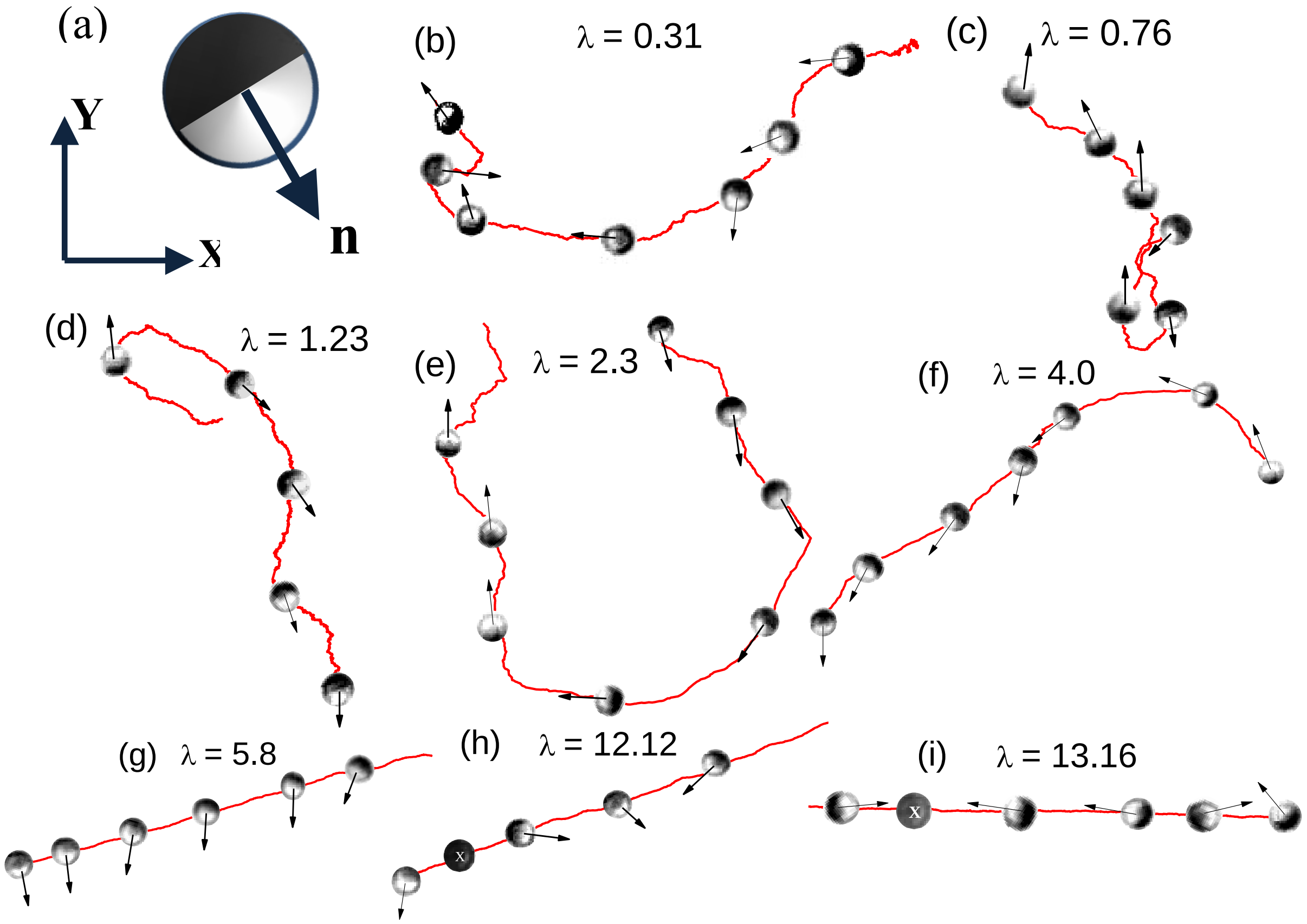

Figure 5: (a) Schematic of the orientation of a Janus particle in X-Y plane as viewed from above (b-i) Representative trajectories of active JPs with snap shots at different times.

Since the direction of propulsion of an active JP depends on orientation, we believe that the variation in the dynamics with $\lambda$, likely reflects the variation in the orientation of JPs. Guided by this underlying reason, we focused on the orientation of the Janus boundary with respect to the flow direction (-X). Figure 5(a) illustrates the notation of the normal to the Janus boundary (**n**), which is useful in the following discussion. In figure 5 (b-i), we show the representative trajectories at different $\lambda$ with snapshots of the JPs at intermittent intervals. Here, we intentionally show the trajectories with high values of $\alpha$, to highlight the role of orientation on the migration along Y. It is evident that at small $\lambda$ (=0.31), the JP orients (**n**) randomly in different directions. At higher values of $\lambda$ (=0.4-5.86), where particles show significant cross migration i.e. $\alpha$ >1, it appears that JP's **n** frequently orients nearly perpendicular to the flow direction especially in the regions where it moves along Y. For both low and moderate $\lambda$, we also note that for all times, JPs display orientation with nearly half of their area to be of the metal patch (dark side), which suggests minimal or no out of plane rotation. At the highest values of $\lambda$ (=12.12,13.16), JPs appear to adopt orientations with varying *Pt* area fractions at regular time intervals, indicating out-of-plane rotation i.e. rolling. This suggests that JPs are forced to mainly drift along the flow with insignificant migration across the streamlines.

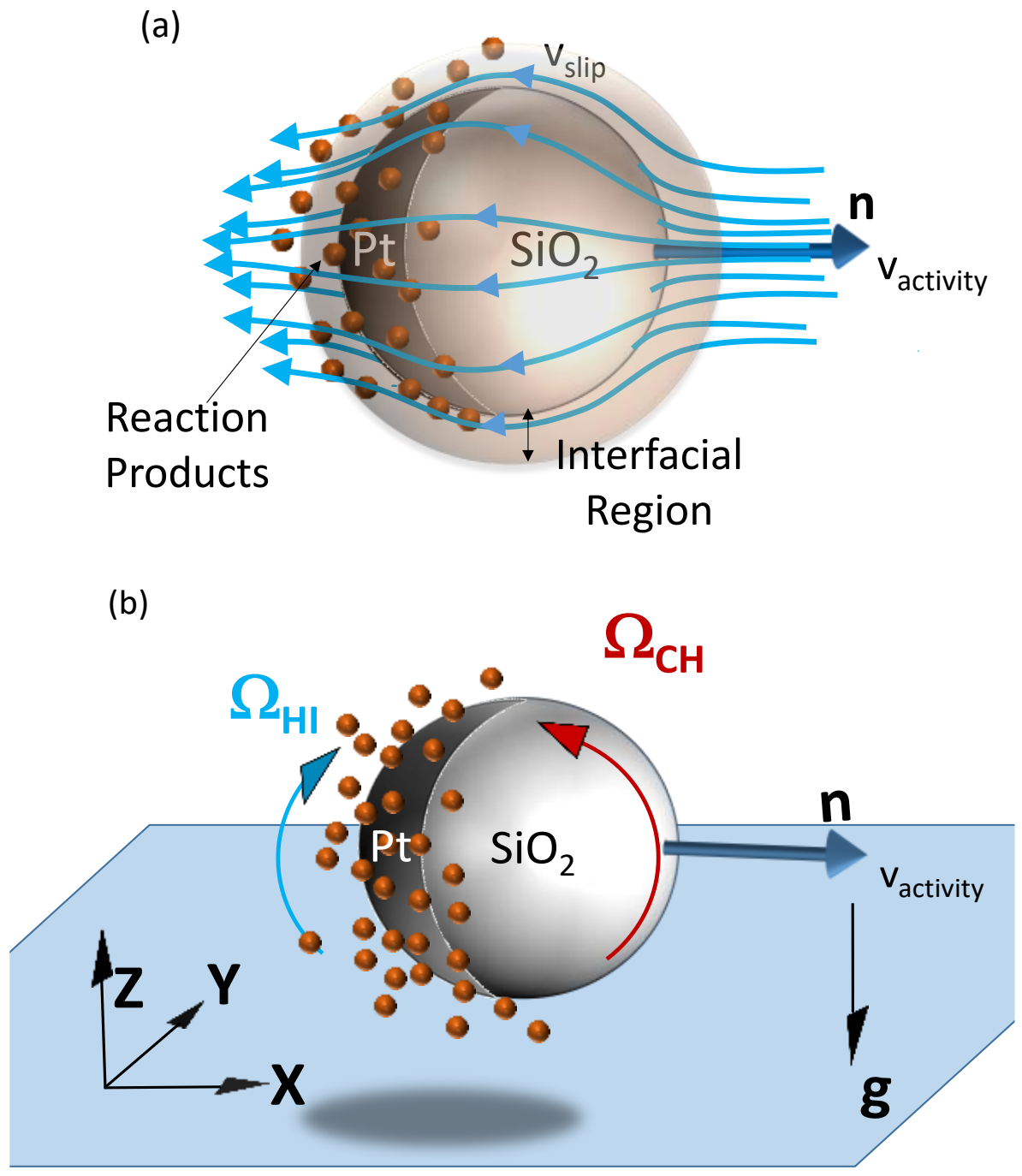

Figure 6: (a) Schematic representation of the mechanism of propulsion for an $SiO_2$-*Pt* Janus particle in the bulk (b) Schematic of the torques acting on an active $SiO_2$-*Pt* Janus particle near a wall. $\Omega_{HI}$ depicts the rotation induced by the hydrodynamic interaction of long-range fluid flow and $\Omega_{CH}$ depicts the rotation due to the chemiosmotic effects.

To understand the influence of flow on the orientation of active Janus particles, we now draw our attention to the presence of a solid wall imposing a no-slip boundary condition in the vicinity of the particle. Schematic shown in figure 6(a), describes the mechanism of propulsion 'self-diffusiophoresis'[10,12] of a $SiO_2$-*Pt* JP suspended in the bulk of stationery $H_2O_2$ aqueous solution. According to the mechanism, the chemiosmotic slip ($v_{slip}$) generated by the non-uniform distribution of the reaction products propels the particle in the opposite direction away from the *Pt* cap. The interfacial flow also generates a torque ($\sim\eta(v_{slip}/a)a^3$) which gets balanced around the particle due to the symmetry of the flow fields, if the particle is in the bulk (i.e. far away from any solid surface). However, when an active JP approaches near a solid surface, new interactions emerge and has been discussed in detail earlier,[20,26,35,39,41] which we summarize here. Firstly, the long-range fluid flow induced by the particle motion gets reflected by the wall generating a hydrodynamic coupling with the particle which tries to rotate the particle in clockwise direction ($\Omega_{HI}$) (see figure 6(b). It has been shown analytically,[20,26,35,41] that depending on the relative strengths of these competing torques, for a given height of the particle from the surface, the active JP typically orients its **n** almost parallel to the surface. In fact, any out-of-plane perturbation in **n** due to Brownian motion is quenched by this resultant activity (hydrodynamic + chemiosmotic) induced restoring rotation ($\Omega_{activity}$). However, this does not hinder the rotation of **n** about an axis perpendicular to the wall i.e. in-plane-rotation. Different experiments have also verified this observation.[26,39,42] In supporting figure 2S we show the snapshot of actively moving JPs at 2 wt% which confirms that almost all the JPs are oriented with **n** along the X-Y plane. Das *et al.*[39] suggested a phenomenological Hookean form for the angular velocity of rotation $\Omega_{activity} \sim v_{activity}/a$, associated to this restoring mechanism. Since, at the particle level Re <<1, we can approximate the restoring torque $|L_{activity}| \sim \eta a^3|\Omega_{activity}|$, where $\eta$ is the bulk water viscosity. On the other hand, as shown in figure 7, external shear flow tries to roll the Janus particle about the vorticity axis (parallel to the plane and perpendicular to the flow i.e. Y) by imposing a characteristic torque of rotation $|L_{shear}| \sim \eta a^3|v_{shear}/a| \sim \eta a^3|v_{p,shear}/a|$. The rolling of the particle is associated with the rotation of **n** about Y axis as well, which can be visualized as $\mathbf{n}_{xz}$ (projection of **n** on the X-Z plane) performing circular motion in the shear X-Z plane, figure 7. The radius of the

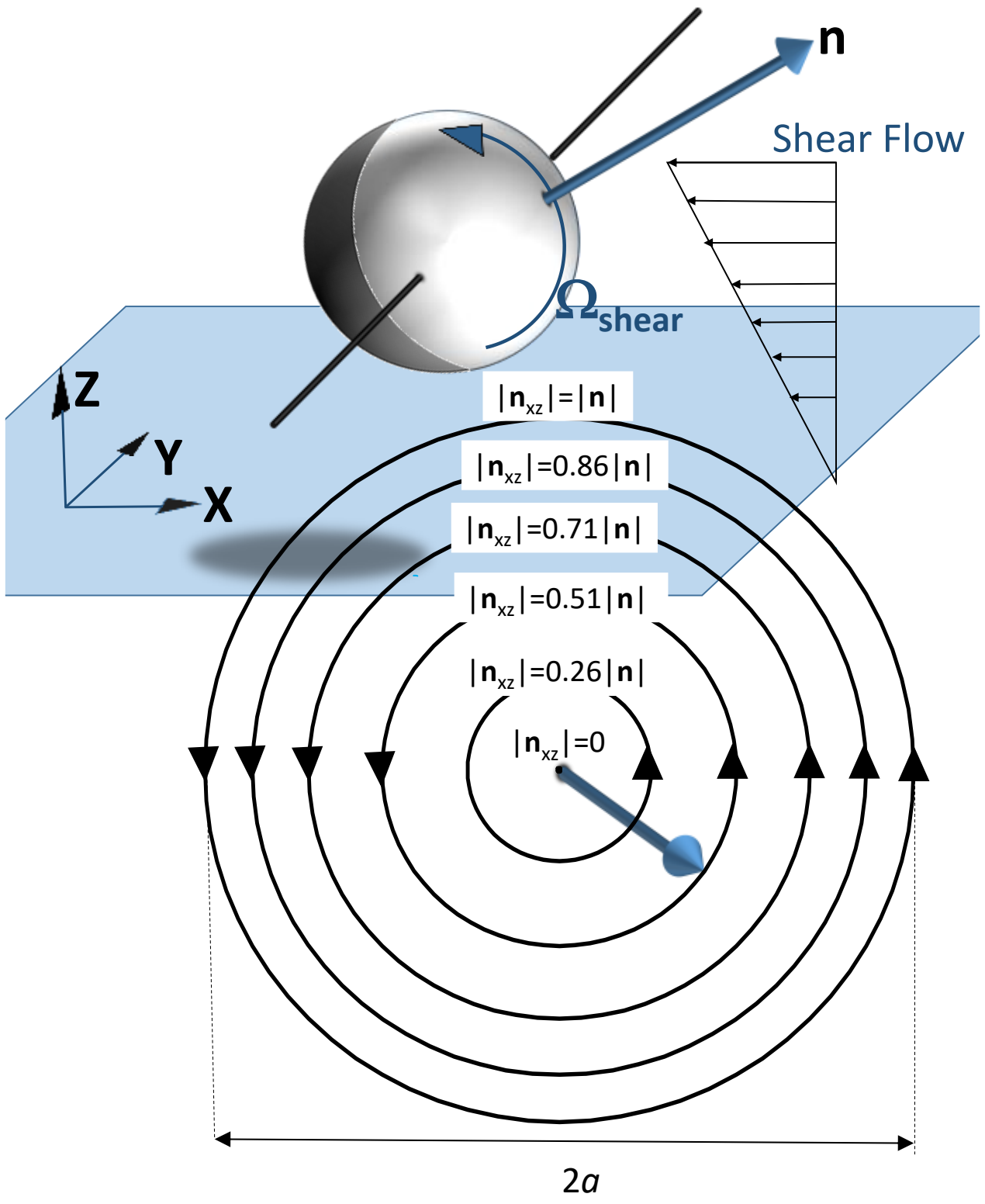

Figure 7: Schematic representation of a passive Janus particle near a solid wall exposed to external shear flow. Inset shows the calculated trajectories of **n** in X-Z plane with varying $|\mathbf{n}_{xz}|$.

circular trajectories $|\mathbf{n}_{xz}|$ is a measure of shear induced out-of-plane rotation of **n**. It is evident from figure 7 that for $|\mathbf{n}_{xz}|$=0 i.e. when **n** is along Y, rolling of Janus particle does not impose any rotation of **n** which increases with $|\mathbf{n}_{xz}|$ with maximum for $|\mathbf{n}_{xz}|=|\mathbf{n}|$ i.e. when **n** is in X-Z plane. If the Janus particle is active, activity will resist the shear induced rotation of **n** about the Y axis and the relative strengths of the competing torques will determine the final orientation of the active JPs. It is noteworthy that the ratio of these two torques $|L_{shear}|/|L_{activity}| \sim |v_{p,shear}|/|v_{activity}|$ which is same as previously defined $\lambda$. In light of our experimental observations and the discussion above, using the cartoons shown in figure 8, we now discuss different regimes of $\lambda$ as following:

At small values of $\lambda$, since $|L_{shear}| << |L_{activity}|$, an attempt to roll the particle by the shear flow is easily resisted by the activity induced restoring torque, such that **n** remains parallel to the planar wall. Also, since $|v_{p,shear}| << |v_{actvity}|$, the in-plane orientation of **n** remains un-biased towards any particular direction (see figure 5(b)). Under these conditions, activity does not drive the active particles preferentially in any particular direction, resulting in $\alpha \sim 1$.

At moderate $\lambda$, since $|L_{shear}| \sim |L_{activity}|$, JPs would be driven towards an orientation where shear flow does not induce any out-of-plane rotation of **n**, to avoid resistance from the activity. From our discussion of figure 7, it is clear that when **n** orients itself along Y i.e. $|\mathbf{n}_{xz}|$=0, rolling of particle does not enforce any out-of-plane rotation of **n**.

Therefore, we hypothesize that this orientation acts like a metastable state for the active JPs in shear flow near a wall. Irrespective of the initial configuration, the orientation of the JPs evolves to this 'biased' orientation through in-plane fluctuations, and that this state remains largely unperturbed by smaller Brownian fluctuations. However, intermittently whenever the fluctuations are larger, particles deviate from that orientation which can be revisited later. For moderate $\lambda$, since $|v_{p,shear}| \sim |v_{actvity}|$, particles with biased orientation preferentially migrate along Y which increases $\alpha$.

achieve that orientation, but since $|v_{p,shear}| >> |v_{actvity}|$, the relative migration in Y remains weaker (lower $\alpha$) compared to the earlier case of moderate $\lambda$. Previous report by Katuri *et al.*[36] demonstrated cross stream migration of active particles in flow for $\lambda$=2.3, 4, 4.67, which is consistent with our proposed criterion for significant cross migration to be noticed.

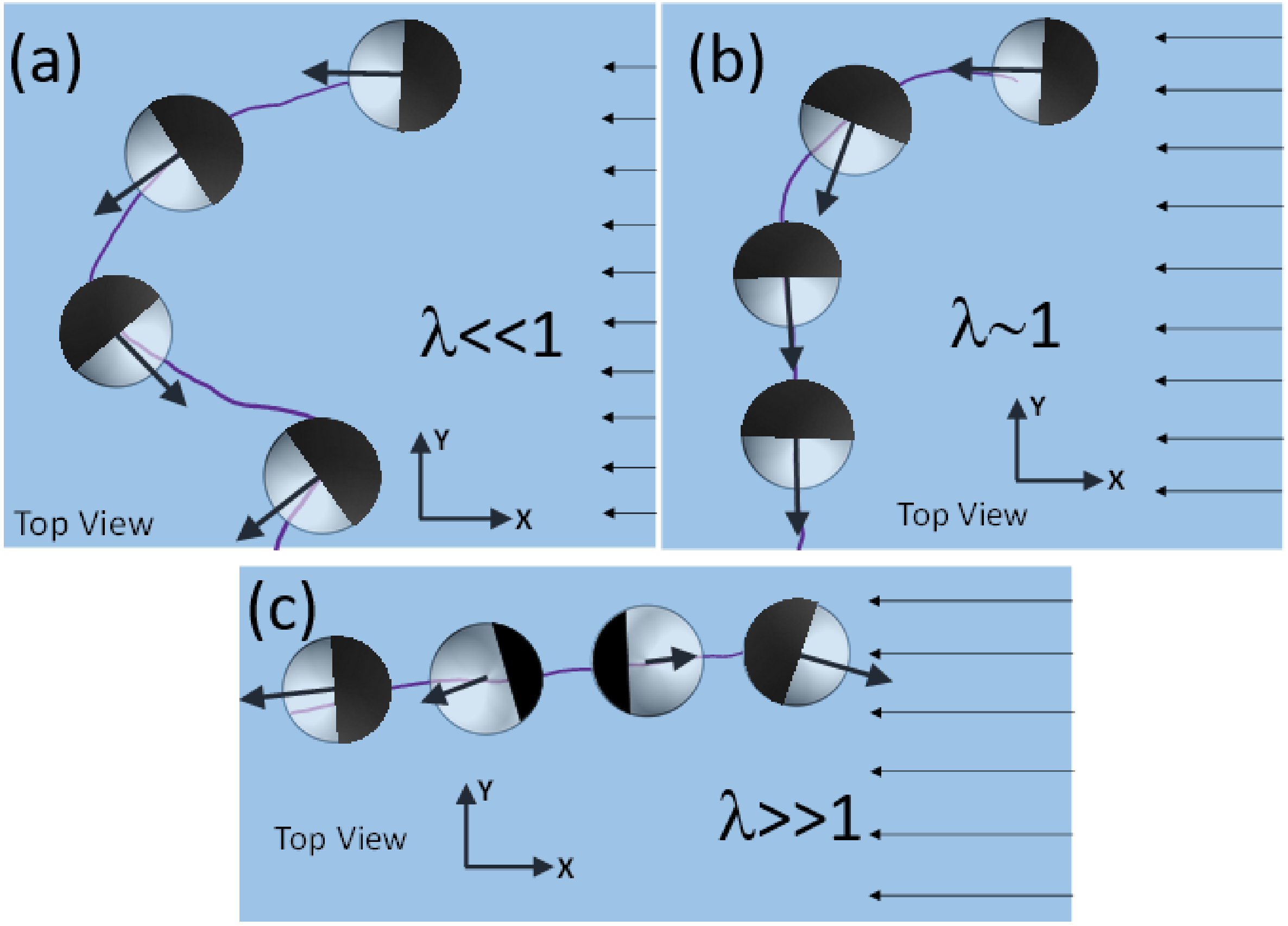

Figure 8: Schematic representation of active $SiO_2$-Pt JPs near passive Janus particle near a solid wall exposed to external shear flow for (a) $\lambda << 1$, active JPs perform random active motion (b) $\lambda \sim 1$, active JPs preferentially migrate perpendicular to the flow direction and (c) $\lambda >> 1$, active JPs drift with the flow.

At high $\lambda$, since $|L_{shear}| >> |L_{activity}|$, the relatively weaker activity induced restoring torque is unable to resist the torque due to shear. This forces the Janus particles to roll and drift with the flow. As shown earlier in figure 5(h,i), we observe snapshots of a JP with **n** showing complimentary azimuthal orientations along with an orientation with *Pt* cap upwards, which confirms the rolling of active JPs.

At high $\lambda$, the faster rolling of JP, outweighs the in-plane rotational motion of the particle which is evident from significantly reduced in-plane rotational diffusivity $D_r$ (see figure 6S). Therefore, it becomes less likely for the particles to encounter the biased orientation with $|\mathbf{n}_{xz}|$=0. A small fraction of JPs, however, may still

## CONCLUSIONS

Through carefully performed experiments, we build upon the previous studies of active motion near a solid boundary. We observe, that the complex interplay between the activity, hydrodynamic interactions with the wall and the shear induced rotation governs the direction of motion of active colloids. At low shear rates, since the shear induced torque is weak, it is easily resisted by the activity induced restoring torque. Hence, JPs orient their normal to the Janus plane (**n**) parallel to the wall boundary and due to the unrestricted in-plane rotation, these JPs perform random active motion with no bias towards any direction. However, at high flow rates, since shear induced torque dominates the restoring torque, particles are forced to roll and move with the flow with little migration perpendicular to the flow. At intermediate conditions, when the external flow strength is comparable to particle's propulsion speed, JPs preferentially favor the orientation where their **n** aligns along the vorticity direction. Under these circumstances, particles migrate significantly perpendicular to the flow direction. We supported these conclusions through our measurements of the anisotropy in the trajectories estimated using a non-dimensional number $\alpha = \frac{\left( \Delta Y_{max} / \Delta X_{max} \right)_{activity+flow}}{\left( \Delta Y_{max} / \Delta X_{max} \right)_{flow}}$ and the in-plane orientation of Janus boundary normal in flow. While the observation of the cross-migration of spherical active Janus particles in flow near a planar wall is not new (credited to Katuri *et al.*), to the best of our knowledge, for the first-time, we carefully delineated the conditions for this phenomenon to be observed. Through our carefully designed experiments, we demonstrated that the active JPs exhibited a continuous transition in their preferred direction of motion governed by a non-dimensional number $\lambda = \frac{| v_{p,shear} |}{| v_{activity} |}$, which estimates the strength of the shear flow relative to the strength of the activity.

In the end, we would like to conclude that our investigation of the active motion of $SiO_2$-Pt JPs in a simple Newtonian medium under shear flow near a planar wall unmasks novel fundamental insights in the area of active colloids especially in the context directing the transport and delivery of functional particles (e.g., active agent-loaded) in flow. Finally, we note that this study suggests a number of directions for future research. For example, it should be possible to gain further control over the active motion of JPs by tuning the orientation of JPs using an elastic media. In addition, it would be interesting to explore the effects of active particles on the dynamics of passive colloids in an external shear flow.

## AUTHOR INFORMATION

### Corresponding Author

* Email: mangalr@iitk.ac.in

## ORCID

Rahul Mangal: 0000-0003-1824-417X

### Author Contributions

The manuscript was written through contributions of all authors. All authors have given approval to the final version of the manuscript.

### Funding Sources

This work is supported by the Science and Engineering Research Board (SERB/CHE/2019077), Department of Science and Technology, India.

### Notes

Authors declare no competing financial interest.

## ACKNOWLEDGMENT

We acknowledge Prof. Ashish Garg and Prof. Shashank Kumar for their assistance with Platinum deposition. We also thank Prof. Indranil Saha Dalal and Prof. Harshwardhan Katkar for useful discussions.

# Supporting Information

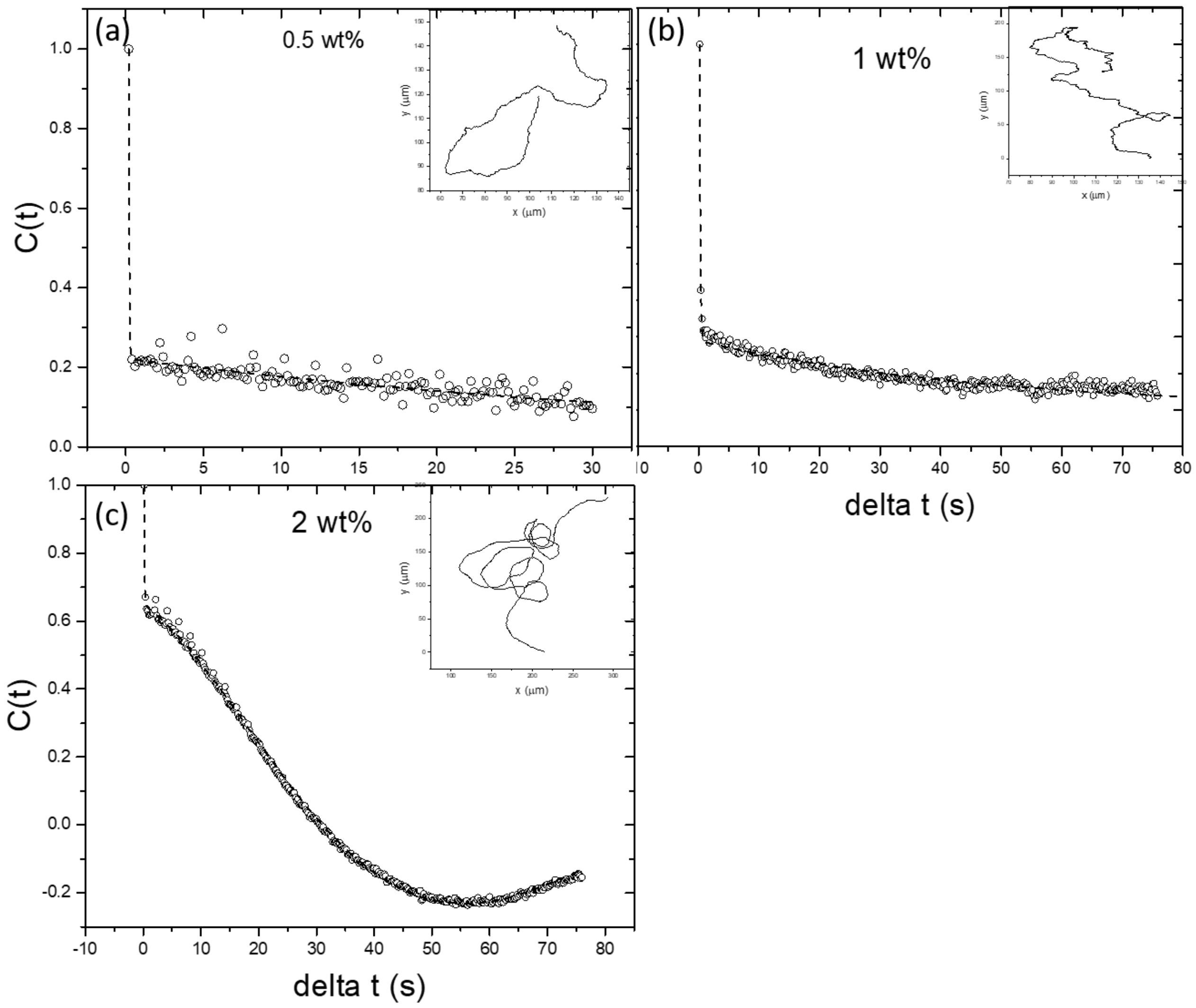

Figure **Error! No text of specified style in document.**1S: Velocity autocorrelation function C(t) vs del t for active particles without flow with (a) 0.5% $H_2O_2$

(b) 1 % $H_2O_2$ (c) 2% $H_2O_2$

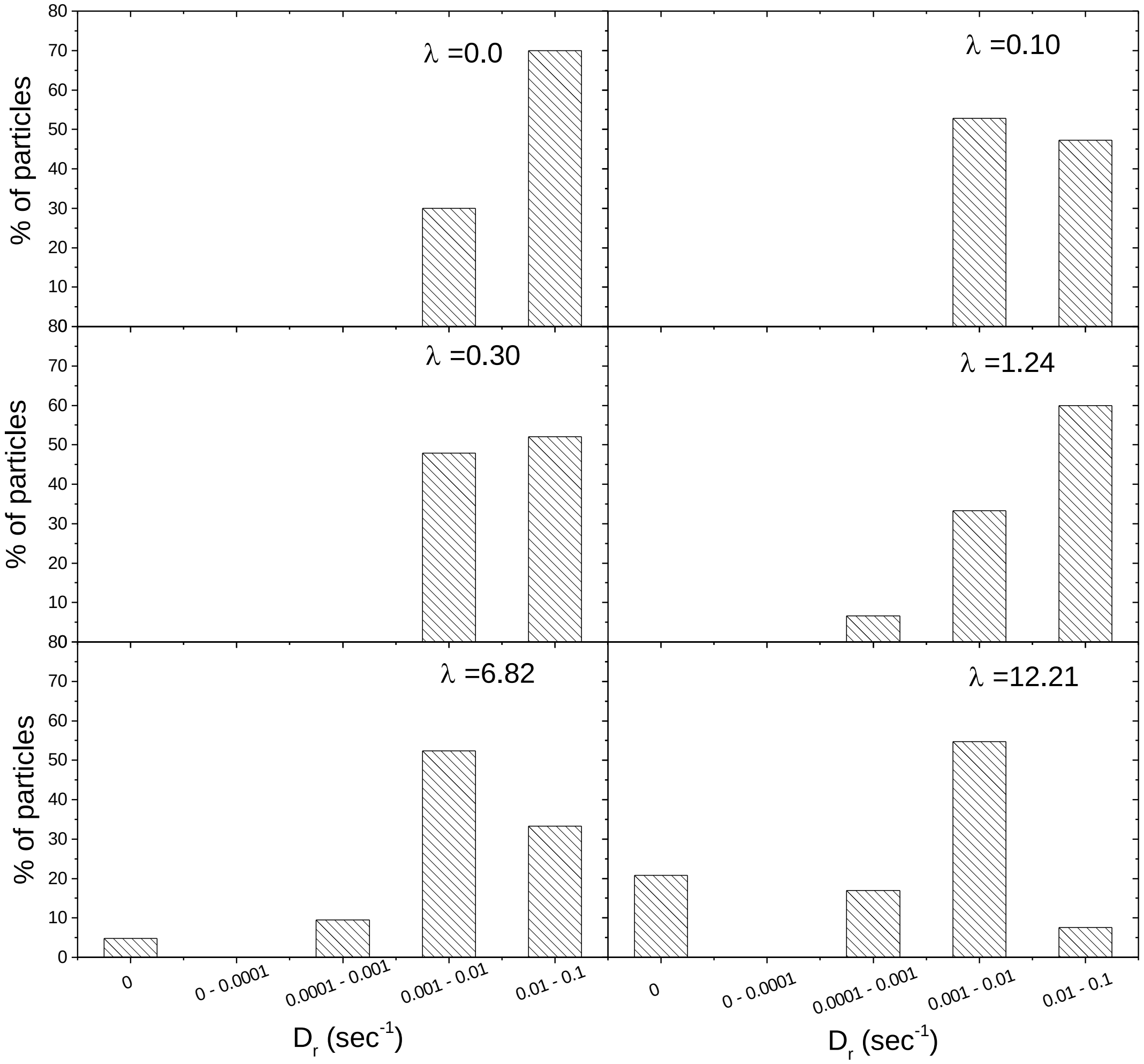

Figure **Error! No text of specified style in document.**2S: Distribution of rotational diffusivity $D_r$ at varying λ.

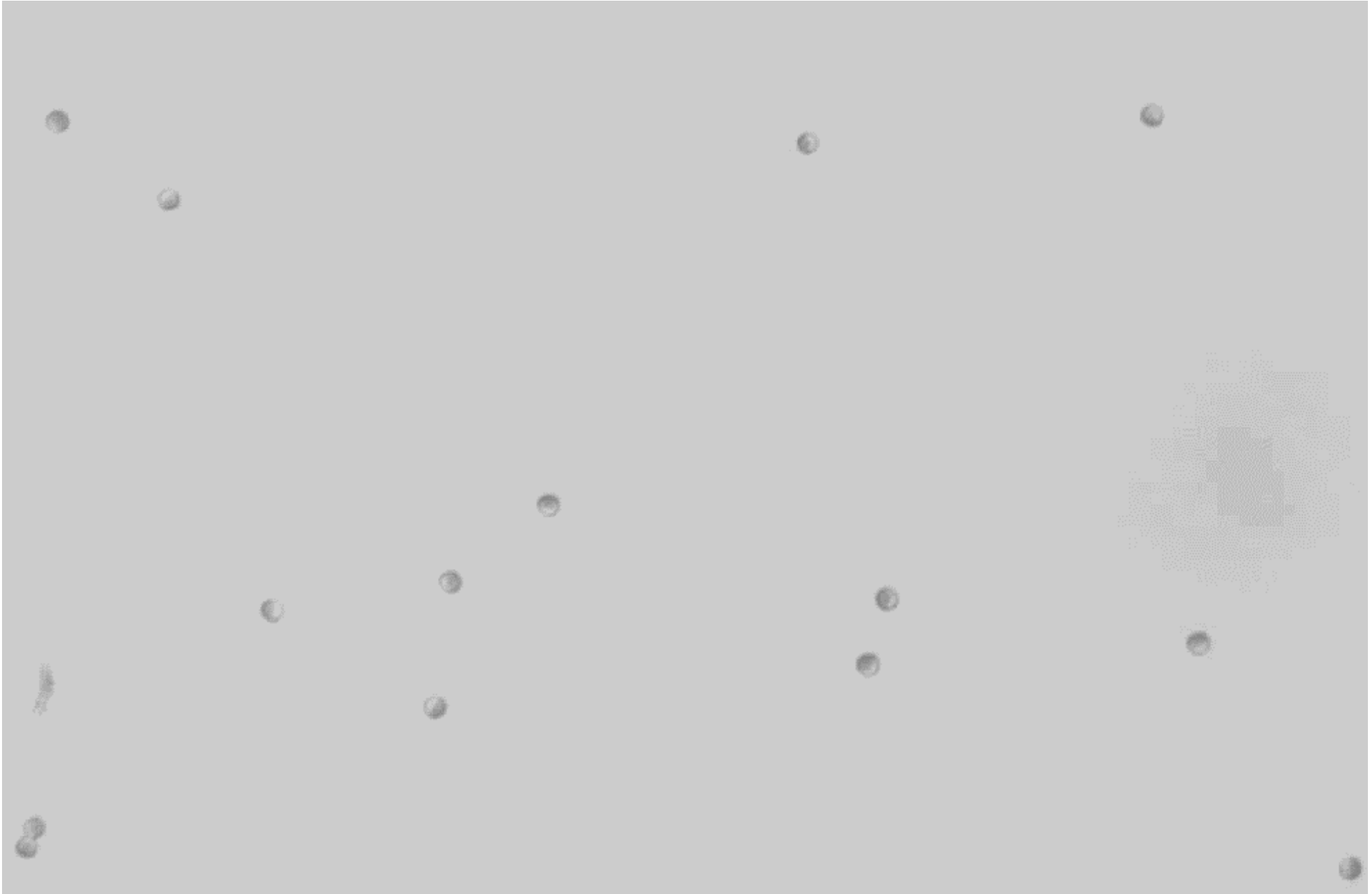

Figure **Error! No text of specified style in document.**3S: Snapshot of active JPs at 2wt% $H_2O_2$